\documentclass[aps,prd,onecolumn,english,showpacs,superscriptaddress,longbibliography]{revtex4-2}

\usepackage{amsmath}
\usepackage{amssymb}
\usepackage{amsfonts}
\usepackage{mathrsfs}
\usepackage[dvipsnames]{xcolor}

\usepackage{amsthm}

\theoremstyle{definition}

\def\XXint#1#2#3{{\setbox0=\hbox{$#1{#2#3}{\int}$}
     \vcenter{\hbox{$#2#3$}}\kern-.5\wd0}}

\usepackage[utf8]{inputenc}
\usepackage[english]{babel}

\usepackage[]{graphicx}
\usepackage[colorlinks=true,plainpages=false]{hyperref}

\newcommand{\beq}{\begin{equation}}
\newcommand{\eeq}{\end{equation}}
\newcommand{\bea}{\begin{eqnarray}}
\newcommand{\eea}{\end{eqnarray}}

\definecolor{green}{rgb}{0.1,.7,0.05}

\begin{document}
 \title{Analytical exploration of the optomechanical attractor diagram and of limit cycles} 

\author{Jorge G. Russo}\email{jorge.russo@icrea.cat}
\affiliation{Instituci\'o Catalana de Recerca i Estudis Avan\c{c}ats (ICREA), \\ Pg.~Lluis Companys, 23, 08010 Barcelona, Spain}\affiliation{Departament de F\' \i sica Cu\' antica i Astrof\'\i sica and Institut de Ci\`encies del Cosmos, \\ Universitat de Barcelona, Mart\'i Franqu\`es, 1, 08028 Barcelona, Spain }
\author{Miguel Tierz}\email{tierz@simis.cn}
%\email{...}
\affiliation{Shanghai Institute for Mathematics and Interdisciplinary Sciences\\ Block A, International Innovation Plaza, No. 657 Songhu Road, Yangpu District,\\ Shanghai, China}
\affiliation{Departamento de An\'alisis Matem\'atico y Matem\'atica Aplicada, Universidad Complutense de Madrid, 28040 Madrid, Spain}

%(上海数学与交叉学科研究院)

\begin{abstract}
We analyse the interplay between mechanical and radiation pressure in an optomechanical cavity system. Our study is based on an analytical evaluation of the infinite Bessel summations involved, which previously had led to a numerical exploration of the so-called attractor diagram. The analytical expressions are then suitable for further asymptotic analysis in opposing regimes of the amplitude, which allows for a characterisation of the diagram in terms of elementary functions. Building on this framework, we investigate the emergence and properties of optomechanical limit cycles beyond the constraints of the resolved sideband approximation. By employing a Fokker-Planck formalism originally developed in the context of laser theory and then used in cavity optomechanics, we describe the quantum regime of these limit cycles, offering a more detailed and unified analytical perspective.

\end{abstract} 
\maketitle
{
  \hypersetup{linkcolor=black}
  \tableofcontents
}

\section{Introduction}

Maxwell predicted that electromagnetic radiation can exert
forces on material objects \cite{maxwell1873treatise}. Typically, this force is negligible,
but it has important effects in specific situations, such as in the tails of comets or during star formation. Optical forces have been used since the 1970s to trap and manipulate small particles and even individual atoms; see \cite{bradshaw2017manipulating} for example, for early references. There is currently great interest in using radiation forces to manipulate the centre of mass motion of mechanical oscillators, ranging from the nanoscopic to the macroscopic (LIGO project) \cite{aspelmeyer2014cavity}.
The phenomenon of damping can be described as a process whereby the motion of a system is slowed down by the application of a force. This force can be either positive or negative, with the former being responsible for cooling the motion and the latter for amplifying small forces \cite{kippenberg2005analysis}.\\

While cooling motion is one of the cornerstones of cavity optomechanics, we focus here on the amplification features. Indeed, the light-induced forces can also result in a negative contribution to the overall damping rate. Initially, this increases the mechanical quality factor (Q) of the mechanical degree of freedom (for example, a cantilever), thereby amplifying the response to any noise source acting on the cantilever. Once the overall ‘damping rate’ becomes negative, which can occur simply by increasing the light power entering the cavity, the system no longer exhibits damping behaviour but instead displays instability \cite{dorsel1983optical,braginsky2001parametric}.\\

More precisely, we begin with the work carried out on a standard optomechanical system \cite{marquardt2006dynamical,marquardt2007quantum}, which involves a high-finesse optical cavity formed by two mirrors: one fixed and the other mounted on a flexible cantilever. The position of the movable mirror, $x$, affects the cavity length, thereby shifting the resonance frequency and modulating the amplitude of the light inside the cavity, $\alpha$. The radiation pressure force, proportional to \(|\alpha|^2\), acts back on the mirror to create a coupled optomechanical system. This nonlinear coupling gives rise to complex dynamical phenomena, including multistability and self-sustained oscillations. For certain parameters, the static force balance has multiple stable points, leading to bistability. The combination of radiation pressure delays and mechanical damping creates dynamical instabilities, causing self-sustained oscillations of the cantilever. The system exhibits a range of stable oscillation amplitudes (dynamic attractors) that depend on the detuning, \(x_0\) and the ratio between the mechanical and radiation pressure \(\Gamma/P\).\\
  
In the next section, we present a methodology for the analytical study of the expression for the radiation pressure in the above described model and then the ensuing attractor diagram that has been numerically outlined in references \cite{marquardt2006dynamical,marquardt2007quantum}. The determination of radiation pressure represents a fundamental problem in optomechanical systems and the force and power input are well-known to be characterised by infinite sums of Bessel functions \cite{marquardt2006dynamical,aspelmeyer2014cavity}, as shown in the next Section. Likewise, as we will see at the end of the Outlook section, these infinite summations are ubiquitous in a wide range of problems in cavity optomechanics and magnomechanics and beyond, including an extremely large number of quantum Floquet problems. This paper will exploit that these sums can be actually computed exactly \cite{russo2024landau,tierz2025analytical} and discuss a number of interesting predictions from the exact formulas, focusing exclusively on the dynamical multistability \cite{marquardt2006dynamical,marquardt2007quantum,ludwig2008optomechanical} and limit cycle problems in cavity optomechanics \cite{lorch2014laser,amitai2017synchronization}.\\

Indeed, optomechanical systems are known to exhibit self-sustaining limit cycles in which the quantum state of the mechanical resonator exhibits non-classical properties \cite{lorch2014laser}. The second part studies such limit cycles. We explain here tentatively some of the basic aspects of \cite{lorch2014laser}, that introduced several simplifications over the usual master equation  \cite{aspelmeyer2014cavity,meystre2013short} describing the optomechanical system. In particular, an effective equation of motion of the Fokker-Planck type was derived in \cite{lorch2014laser} for the motion of the mirror, based on the assumption that the cavity dynamics adiabatically follows the mechanical oscillator (assumption valid when the cavity decay rate $\kappa$ is larger than the characteristic coupling strength of the oscillator and the cavity mode). Our analytical findings apply to all cases presented in \cite{lorch2014laser,amitai2017synchronization}. It should be stressed that other works at the same time carried out similar analysis, for example \cite{rodrigues2010amplitude,armour2012quantum}, using a truncated Wigner function.\\

One of the applications of the effective equation of motion is the subsequent description of optomechanical limit cycles in the quantum regime. The objective of this paper as it concerns to limit cycles is to characterise the drift and diffusion coefficients in \cite{lorch2014laser} again through the utilisation of exact evaluations of Bessel summations, previously employed in \cite{russo2024landau,tierz2025analytical}. The results will be compared with those presented in \cite{lorch2014laser}, which did not consider the exact summation of the formulas. This led the authors of that work to consider a drive setting close to resonance, specifically in the sideband-resolved regime with a detuning close to the resonance. This resonance implies that only the first terms in the sums need to be retained. We investigate the differences and further possibilities afforded by considering the summation in full, analytically. For example, some limit cycles that  appear in \cite{lorch2014laser} turn out to be  an artifact of the approximation as they disappear as solutions in the  exact formula.  We will also specifically examine the implications of the exact summation of the diffusion coefficient for the Wigner function.

%The summations discussed here could be considered in many of such works; however, for the sake of clarity, we will focus on the results presented in reference \cite{lorch2014laser} besides on some well-known results for the radiation pressure presented in references \cite{marquardt2006dynamical,aspelmeyer2014cavity}.

\section{Radiation pressure: parametric instability and Cooling}

In the last decades the classical theory of the optomechanical parametric 
instability produced by radiation pressure backaction has been extensively studied in a large number of works, starting with the influential experimental works \cite{carmon2005temporal,kippenberg2005analysis}. We shall follow the very well-known ensuing theoretical analysis of \cite{marquardt2006dynamical,ludwig2008optomechanical}, where a numerical analysis of the analytical expressions of the model were carried out, culminating in a rather pervasive attractor diagram, that has also been portrayed in many reviews and works afterwards \cite{kubala2009optomechanics}.\\

\begin{figure}[h!]
 \centering
 \includegraphics[width=0.42\textwidth]{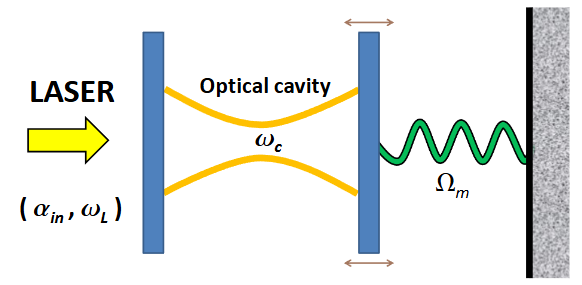}
  \caption
 {Generic optomechanical setting with detuning \(\Delta = \omega_L - \omega_c\) and $G = \partial \omega_c / \partial x$ (frequency shift of the cavity resonance per unit mechanical displacement). The classical coupled equations of motion are given below in \eqref{alpha} and \eqref{x}.}
 \label{optodiag}
 \end{figure}

The model studied in these works describe the nonlinear dynamics of an optical cavity in which one mirror is mounted on a flexible mechanical element (a cantilever). The
cavity is driven by a laser, and the radiation pressure exerted by the
cavity light couples the mechanical motion of the mirror to the
intracavity optical field. A consequence of our analytical formulas obtained below will be a quantitative analysis of this attractor diagram for certain regions of the parameter space. These results apply to rather generic optomechanical systems, consisting of a driven
optical cavity and a movable mirror attached to an oscillator, see the fig. \ref{optodiag}.\\

Writing $\alpha(t)$ for the slowly varying complex amplitude of the intracavity optical field, which is the classical equivalent of the cavity photon field operator $\alpha(t)=\left\langle \widehat{a}(t)\right\rangle$, the coupled classical nonlinear dynamics of the system are governed by the following coupled equations of motion for $\alpha(t)$ and for  the mechanical displacement of the oscillator from its equilibrium position $x(t)=\left\langle \widehat{x}(t)\right\rangle$ \cite{marquardt2006dynamical,ludwig2008optomechanical,aspelmeyer2014cavity}:
%\begin{align}
 %   \dot{\alpha} = \left[i x - \frac{1}{2}\right] \alpha + \frac{1}{2}\\
%\ddot{x} = P |\alpha|^2 - \omega_0^2 (x - x_0) - \Gamma \dot{x}
%\end{align}
\begin{align}
\dot{\alpha} = -\frac{\kappa}{2} \alpha + i(\Delta + G x) \alpha + \sqrt{\kappa_{ex}} \alpha_{in} , \label{alpha} \\
m \ddot{x} = -m \Omega_m^2 x - m \Gamma_m \dot{x} + \hbar G |\alpha|^2 + F_{ext} \label{x},
\end{align}
where $\kappa$ denotes the total energy decay rate of the optical cavity mode and describes the rate at which energy leaks out of the optical cavity. The parameter $\Delta$ is the detuning of the driving laser frequency \(\omega_L\) from the bare cavity resonance frequency \(\omega_c\), defined as \(\Delta = \omega_L - \omega_c\). $G$ is the optomechanical coupling strength describing the frequency shift of the cavity resonance per unit mechanical displacement and \(\alpha_{in}\) is the input field amplitude (drive) of the laser incident on the cavity. The effective mass of the mechanical oscillator is $m$ and the intrinsic mechanical resonance angular frequency is $\Omega_m$. The mechanical energy decay rate (damping rate), describing mechanical losses is denoted by $\Gamma_m$. Finally, $|\alpha|^2$ denotes the intracavity photon number, i.e. the intensity of the optical field inside the cavity and $F_{ext}(t)$ denotes any other external force acting on the mechanical oscillator (e.g. thermal noise or applied forces). We adopted the notation of the review \cite{aspelmeyer2014cavity}.\\

We have adopted the notation in \cite{aspelmeyer2014cavity}, also for the Ansatz below and the ensuing solution, described below. Notice that in the original works \cite{marquardt2006dynamical,ludwig2008optomechanical} the parameters are presented in a seemingly different yet equivalent manner.\\

These coupled equations capture the dynamics of cavity optomechanics in the
classical limit, describing how light inside the cavity influences the motion of the mechanical
oscillator via radiation pressure and how the oscillator, in turn, shifts the cavity resonance
frequency and affects the optical field. More precisely, in \eqref{alpha} the cavity field amplitude \(\alpha\) decays
at a rate \(\kappa/2\), acquires a phase shift depending on the detuning \(\Delta\) and mechanical displacement via \(G x\), and is
driven by the external input field \(\alpha_{in}\) through the coupling rate \(\kappa_{ex}\) and in \eqref{x} the mechanical oscillator behaves like a damped harmonic oscillator with mass \(m\), resonance frequency \(\Omega_m\), damping \(\Gamma_m\), and is driven by the radiation pressure force \(\hbar G |\alpha|^2\) (force from the intracavity photons) plus any other external force \(F_{ext}\).\\

The Ansatz for the self-induced oscillations is \cite{marquardt2006dynamical}
\begin{equation}
x(t)=\bar{x}+A\cos(\Omega_{{\rm m}}t)\ ,
\label{eq:Ansatz}
\end{equation}
The radiation pressure force $F(t)$ is deduced by solving the classical equation for the light field amplitude, which is
\begin{equation}
\dot{\alpha}=-\frac{\kappa}{2}(\alpha-\alpha_{{\rm max}})+i(\Delta+Gx(t))\alpha\,,
\end{equation}
where $\alpha_{{\rm max}}$ is defined to be the amplitude reached
inside the cavity right at resonance,
we have $\alpha_{{\rm max}}=2\alpha_{{\rm in}}\sqrt{\kappa_{{\rm ex}}}/\kappa$. \\ The time-averaged
power input due to this force is $\left\langle F\dot{x}\right\rangle $.
After inserting the ansatz \eqref{eq:Ansatz},
the solution for the optical degrees of freedom  can be obtained as a Fourier series
$\alpha(t)=e^{i\varphi(t)}\sum_{n}\alpha_{n}e^{in\Omega_{{\rm m}}t}$,
with coefficients \cite{aspelmeyer2014cavity,marquardt2006dynamical}
\begin{equation}
\alpha_{n}=\frac{\alpha_{{\rm max}}}{2}\frac{J_{n}(-\frac{GA}{\Omega_{{\rm m}}})}{in\frac{\Omega_{{\rm m}}}{\kappa}+\frac{1}{2}-i(G\bar{x}+\Delta)/\kappa}\,,
\end{equation}
where  the global
phase is $\varphi(t)=(GA/\Omega_{{\rm m}})\sin(\Omega_{{\rm m}}t)$.
Now the force $F(t)=\hbar G\left|\alpha(t)\right|^{2}$ and the time
averages $\left\langle F\dot{x}\right\rangle $ and $\left\langle F\right\rangle $
can be calculated. One has \cite{marquardt2006dynamical,aspelmeyer2014cavity}
\begin{equation}
\left\langle \left|\alpha(t)\right|^{2}\right\rangle =\sum_{n}\left|\alpha_{n}\right|^{2},    
\end{equation}
and
\begin{equation}
\left\langle \left|\alpha(t)\right|^{2}\dot{x}\right\rangle =A\Omega_{{\rm m}}{\rm Im}\sum_{n}\alpha_{n}^{*}\alpha_{n+1}.
\label{hhy}
\end{equation}

\subsection{Analytical characterization}

These series have always been dealt with numerically. Following the same analytical results employed while studying superconducting qubits \cite{russo2024landau} (see Appendix B for a summary), we can now give their full analytical form. For the time-averaged force we obtain
\begin{equation}
\left\langle F(t)\right\rangle=\hbar G\left\langle \left|\alpha(t)\right|^{2}\right\rangle =\hbar G\frac{\kappa \big|\alpha_{{\rm max}}\big|^2}{2\Omega_{{\rm m}}}   
{\rm Im}\left( \frac{\pi}{\sin(\pi\rho) }\, J_\rho (X ) J_{-\rho} (X)\right) \ ,
\end{equation}
where
\begin{equation}
\qquad X\equiv  -\frac{GA}{\Omega_{{\rm m}}}\ ,\ \ \rho \equiv \frac{1}{\Omega_m}(G\bar x+\Delta+\frac{i \kappa }{2} )\ .
\end{equation}

To evaluate \eqref{hhy}, we first write
\begin{equation}
\alpha_{n}^{*}\alpha_{n+1}=\frac{\kappa^2\big|\alpha_{{\rm max}}\big|^2}{4\Omega_{{\rm m}}^2}\ \frac{J_n(X)J_{n+1}(X)}{(n-r+ic)(n+1-r-ic)}   \ ,\qquad r\equiv \frac{1}{\Omega_{{\rm m}}}(G\bar x+\Delta)\ ,\ \ c\equiv\frac{\kappa }{2\Omega_{{\rm m}}}\ \label{c}.
\end{equation}
Thus
\begin{equation}
\alpha_{n}^{*}\alpha_{n+1}=\frac{\kappa^2\big|\alpha_{{\rm max}}\big|^2}{4\Omega_{{\rm m}}^2}\ \frac{J_n(X)J_{n+1}(X)}{1-2i c} \left( \frac{1}{n-r+ic}-\frac{1}{n+1-r-ic} \right)\ .
\end{equation}
We can now apply the exact summation formula.
We find, for the time-averaged power input due to the radiation force
\begin{equation}
\left\langle F\dot{x}\right\rangle=\left\langle \left|\alpha(t)\right|^{2}\dot{x}\right\rangle = -A
\frac{\kappa^2\big|\alpha_{{\rm max}}\big|^2}{4\Omega_{{\rm m}}}
{\rm Im}\left[\frac{1}{1-2i c} \left(\frac{\pi}{\sin(\pi\nu)} J_\nu(X) J_{1-\nu}(X) + \frac{\pi}{\sin(\pi\nu^*)} J_{1+\nu^*}(X) J_{-\nu^*}(X)\right)\right] .
\label{hhyy}
\end{equation}
with $\nu\equiv -r+ic$.

Similarly as in the previous examples, one can give the asymptotic formulas at $|X|\gg 1$, $|X|\gg|\nu|^2$ using \eqref{asymJ}.
In particular, we find for the force average:
\begin{equation}
    \left\langle F(t)\right\rangle=\hbar \frac{\kappa \big|\alpha_{{\rm max}}\big|^2}{A} b 
    \left(\cosh(\frac{\pi\kappa}{2\Omega_{{\rm m}}} )- \cos(\pi r)\sin(\frac{2GA}{\Omega_{{\rm m}}})\right)\ , \qquad \frac{GA}{\Omega_{{\rm m}}}\gg 1 , 
\end{equation}
with
\begin{equation}
  b \equiv\frac{\sinh(\frac{\pi \kappa}{2\Omega_{{\rm m}}})}{\cosh(\frac{\pi\kappa}{\Omega_{{\rm m}}} )-\cos(2\pi r) } \ ,
\end{equation}
and
\begin{equation}
\left\langle F\dot{x}\right\rangle=\left\langle \left|\alpha(t)\right|^{2}\dot{x}\right\rangle = A
\frac{\kappa^2\big|\alpha_{{\rm max}}\big|^2}{4\Omega_{{\rm m}}} \frac{8 c \cosh (\pi  c) \sin (\pi  r) \cos (2 X)}{\left(4 c^2+1\right) X (\cos (2 \pi  r)-\cosh (2 \pi 
   c))}\label{underpressure}
\end{equation}
From these formulas one can see that the functional dependence of the time-averaged force 
satisfies properties for special 
values of the parameters:

\begin{itemize}

\item When $r=n+\frac12$, with integer $n$, one has $\cos(\pi r)=0$
so the coefficient of the $\sin(\frac{2GA}{\Omega_{{\rm m}}})$ vanishes in $\langle F\rangle$. Therefore oscillations of $\langle F\rangle$ as a function of $A$ are strongly suppressed at large $A$ ($\langle F\rangle$ then has smaller oscillations of $O(1/A^2)$ originating from subleading corrections to the asymptotic formula \eqref{asymJ} of the Bessel function).

\item $\langle F\rangle$ is a periodic function of $r$ with periodicity $r\to r+2$. However,  at the special values $2X=2GA/\Omega_n=\pi n$, with integer $n$, one has $\sin(\frac{2GA}{\Omega_{{\rm m}}})=0$ and the periodicity of $\langle F\rangle$ changes to $r\to r+1$.

\item For small $\kappa$, and $r$ generic, one has $ \langle F\rangle \propto \kappa^2$. However, at
 integer values of $r$,   $\langle F\rangle$ exhibits  resonance behavior, with peaks values given by
\begin{equation}
    \langle F\rangle\bigg|_{r=n}=\frac{|\alpha_{\rm max}|^2\Omega_m}{\pi A}\left(1-(-1)^n\sin(\frac{2GA}{\Omega_{{\rm m}}})\right) \ .
\end{equation}
Thus, for integer values of $r$, the small factor $\kappa^2$ cancels out against a factor $1/\kappa^2$ originating from the resonance. 
A similar resonance behavior is seen in the exact formula \eqref{hhyy} for $\langle F\dot{x} \rangle$. For generic $r$ is $O(\kappa^3)$ but at integer $r$ becomes $O(\kappa)$. 
%On the other hand, the large $A$ asymptotic formula \eqref{underpressure} for $\langle F\dot{x} \rangle$ vanishes at integer $r$. The contribution comes from subleading corrections to the he asymptotic formula \eqref{asymJ} of the Bessel function,
%the actual large $A$ behavior at integer $r$ being  $ \langle F\dot{x} \rangle \bigg|_{r=n}=O(1/A)$.

\end{itemize}

The first two properties arise exclusively as a result of the exact resummation carried out in the present study.
It should be feasible to experimentally test these remarkable predicted features.

\subsection{Parametric instability, amplification and cooling: The attractor diagram}

The condition of force balance determines the average position of the oscillator, thereby yielding an implicit equation for $\bar{x}$ \cite{marquardt2006dynamical,ludwig2008optomechanical,aspelmeyer2014cavity}, \begin{equation}
\langle\ddot{x}\rangle\equiv0\quad\Leftrightarrow\quad m\omega_{M}^{2}\bar{x}=\left\langle F_{{\rm rad}}\right\rangle =\hbar G\langle\left|\alpha(t)\right|^{2}\rangle. \label{eq:forcebalance}\end{equation} The average radiation force, $\langle F_{\text{rad}}\rangle$ is a function of the parameters $\bar{x}$ and $A$.

The balance between the mechanical power gain due to the light-induced
force, $P_{{\rm rad}}=\left\langle F_{{\rm rad}}\dot{x}\right\rangle $,
and the frictional loss $P_{{\rm fric}}=\Gamma_{M}\left\langle \dot{x}^{2}\right\rangle $
follows from \cite{marquardt2006dynamical,ludwig2008optomechanical} \begin{equation}
\langle\dot{x}\ddot{x}\rangle\equiv0\quad\Leftrightarrow\quad\langle F_{\text{rad}}\dot{x}\rangle=\Gamma_{M}\langle\dot{x}^{2}\rangle.\label{eq:powerbalance}\end{equation}
For each value of the oscillation amplitude $A$ the plot of the
ratio between radiation power input and friction loss, $P_{{\rm rad}}/P_{{\rm fric}}=\langle F_{\text{rad}}\dot{x}\rangle/(\Gamma_{M}\langle\dot{x}^{2}\rangle)$ is analyzed in \cite{marquardt2006dynamical,ludwig2008optomechanical}. Therefore, the study of the ratio has been subjected to graphical and numerical analysis as has been also reviewed on numerous occasions after \cite{marquardt2006dynamical,ludwig2008optomechanical}. The power balance is fulfilled if the ratio is 1, whereas if the power input into the cantilever by radiation pressure is greater than frictional losses (i.e., for a ratio greater than 1), the amplitude of oscillations will increase, otherwise it will decrease. Thus, stable solutions (dynamical attractors) are given by decrements in the ratio while increasing oscillation amplitude (energy).

\medskip

It is now possible to study these features analytically in different regimes, for which simplified analytical expressions can be obtained by either asymptotic or power series expansions of the summed expressions. Experimental exploration of the optomechanical attractor diagram and its dynamics was carried out in \cite{buters2015experimental}.

\subsubsection{Large driving amplitudes and resolved/unresolved sideband limit}
 
The expression \eqref{underpressure} describes the setting of large amplitudes. The region is the outer parabola given by $|X|\gg|\nu|^2$ in the $P_{{\rm rad}}=0$ plane of the attractor diagram. After the identification $X\equiv  -\frac{GA}{\Omega_{{\rm m}}}$, \eqref{underpressure} reads 
\begin{equation}
\left\langle F\dot{x}\right\rangle=\left\langle \left|\alpha(t)\right|^{2}\dot{x}\right\rangle = -
\frac{\kappa^2\big|\alpha_{{\rm max}}\big|^2}{4G} \frac{8 c \cosh (\pi  c) \sin (\pi  r) \cos (2 X)}{\left(4 c^2+1\right)  (\cos (2 \pi  r)-\cosh (2 \pi 
   c))}, \label{pressurelarge}
\end{equation} 
which shows that the dependence with the amplitude $A$ is oscillating but with no power dependence on $A$. This may be compared with  the friction pressure $P_{{\rm fric}}=\Gamma_{M}\left\langle \dot{x}^{2}\right\rangle $, which is always quadratic (see \eqref{eq:Ansatz}). Therefore, for very large amplitudes mechanical friction pressure will always dominate and the ratio goes to $0$, as the attractor diagram also clearly shows (we refer the reader to Figure 2 of \cite{ludwig2008optomechanical}, paying particular attention to the floor of the plot region, which holds for all values of detuning at higher energies). Intermediate amplitudes are still well described by the asymptotic formula \eqref{pressurelarge} if $c$ is not too large. Notice also that, in general, for very large amplitudes, the starting Ansatz itself \eqref{eq:Ansatz} is not correct anyway \cite{marquardt2006dynamical}.\\

\medskip

In \cite{marquardt2006dynamical,ludwig2008optomechanical}, it is observed that for sufficiently strong driving, self-induced oscillations appear around integer multiples of the cantilever frequency, $\Delta\approx n\omega_{M}$. Notice that our solution is generically oscillatory both in the detuning $r$ and the amplitude $X$ (the only dependence on the amplitude) for all of its domain of validity. The denominator term $(\cos (2 \pi  r)-\cosh (2 \pi 
   c))$ is always negative and hence the overall sign depends on the $\sin(\pi r)$ and $\cos(2X)$ terms. \\

   An important parameter in optomechanical cavities is the sideband resolution ratio, 
defined as the ratio of the mechanical resonance frequency to the linewidth of the optical mode \cite{ludwig2008optomechanical}. That is, $1/c$ where $c$ was given in \eqref{c}. This parameter determines whether the optical mode effectively interacts with one or both sidebands of the mechanical mode. The former occurs for $c\sim 1$  and is typically referred to as the resolved sideband regime, while the latter occurs for an optical linewidth comparable to or greater than the mechanical resonance frequency and is referred to as the unresolved sideband regime (or bad cavity limit \cite{cirac1995laser}, $c\gg 1$). Sideband resolution is a measure of the dynamic interactions between the optical and mechanical degrees of freedom. 

\medskip

The attractor diagram in \cite{marquardt2006dynamical,ludwig2008optomechanical} has been studied for the resolved sideband regime. This regime is of interest because it is crucial for ground-state cooling \cite{marquardt2007quantum,wilson2007theory}, and $c=1/20$ has already been realised experimentally in \cite{schliesser2008resolved}. Using \eqref{pressurelarge}, for $r=(2m+1)/4$ and $m=0,1,2,...$ the two sideband resolution limits give
(for simplicity we write the case $m=0$):
\begin{eqnarray*}
\left\langle F\dot{x}\right\rangle  &\simeq &\frac{\kappa ^{2}\big|\alpha _{%
\mathrm{max}}\big|^{2}}{4G}\frac{\sqrt{2}e^{-\pi c}\cos (2X)}{c}\quad \text{ for }%
c\gg 1, \\
\left\langle F\dot{x}\right\rangle  &\simeq &\frac{\kappa ^{2}\big|\alpha _{%
\mathrm{max}}\big|^{2}}{4G}4\sqrt{2}c\cos (2X)\quad\text{ for }c\ll 1.
\end{eqnarray*}%
The functional dependence on $c$ and $X$
remains unchanged for those values of $r$
 that do not lead to vanishing pressure --that is, for non-integer values of $r$.
 For example, for $r=1/2$, then:%
\begin{eqnarray*}
\left\langle F\dot{x}\right\rangle  &\simeq &\frac{\kappa ^{2}\big|\alpha _{%
\mathrm{max}}\big|^{2}}{4G}\frac{2e^{-\pi c}\cos (2X)}{c}\quad \text{ for }c\gg 1,
\\
\left\langle F\dot{x}\right\rangle  &\simeq &\frac{\kappa ^{2}\big|\alpha _{%
\mathrm{max}}\big|^{2}}{4G}4c\cos (2X)\quad \text{ for }c\ll 1.
\end{eqnarray*}

\subsubsection{Small driving amplitudes}

A good approximation to steady-state properties of the system can be obtained for small driving amplitudes $X<<1$, normally valid for weak optomechanical coupling. In this way, we explore another region of the attractor diagram \cite{marquardt2006dynamical,ludwig2008optomechanical}, namely the one near the origin of the energy and detuning plane. We first can show that the radiation pressure is quadratic in the amplitude $X$. Since it is also quadratic in the amplitude for the friction this leads to a comparison of the two corresponding coefficients.\\

It follows immediately by having the Bessel function evaluation \eqref{hhyy}. Using \cite{watson1922treatise} (5.41, Eq1) we have:%
\[
J_{\nu }(x)J_{\mu }(x)=\sum_{m=0}^{\infty }\frac{(-1)^{m}\left( \frac{x}{2}%
\right) ^{\mu +\nu +2m}\left( \mu +\nu +m+1\right) _{m}}{m!\Gamma \left( \nu
+m+1\right) \Gamma \left( \mu +m+1\right) },
\]%
where $(n)_{m}$ denotes the rising factorial (or  Pochhammer symbol). At first order, it is linear in $x$ and that makes the radiation pressure
quadratic in the amplitude, for small amplitudes as expected. We have at
first order:%
\begin{eqnarray*}
\frac{\pi }{\sin \pi \nu }J_{\nu }(x)J_{1-\nu }(x) &\approx &\frac{x}{2\nu
\left( 1-\nu \right) }, \\
\frac{\pi }{\sin \pi \nu ^{\ast }}J_{1+\nu ^{\ast }}(x)J_{-\nu ^{\ast }}(x)
&\approx &\frac{x}{2\nu ^{\ast }\left( 1+\nu ^{\ast }\right) }.
\end{eqnarray*}%
Using these together with the expression obtained for the radiation
pressure, recalling also that $\nu \equiv -r+ic$, we then have 
\begin{eqnarray*}
\left\langle F\dot{x}\right\rangle  &\approx &-A\frac{\kappa ^{2}\big|\alpha
_{\mathrm{max}}\big|^{2}}{4\Omega _{\mathrm{m}}}\mathrm{Im}\left[ \frac{1}{%
1-2ic}\left( \frac{X}{2\nu \left( 1-\nu \right) }+\frac{X}{2\nu ^{\ast
}\left( 1+\nu ^{\ast }\right) }\right) \right]  \\
&=&-GA^{2}\frac{\kappa ^{2}\big|\alpha _{\mathrm{max}}\big|^{2}}{%
2\Omega _{\mathrm{m}}^{2}}\frac{cr}{\left( r^{2}+c^{2}\right)
(c^{2}+(r-1)^{2})(c^{2}+(r+1)^{2})}
\end{eqnarray*}%
Therefore, for stability, after dividing by the friction
radiation $P_{fric}=\pi A^{2}\Omega _{\mathrm{m}}$, the stability criteria
is that%
\begin{equation}
f(r)\equiv
\frac{\left\langle F\dot{x}\right\rangle}{P_{fric}}
=
-2G\frac{\big|\alpha _{\mathrm{max}}\big|^{2}}{\pi 
\Omega_{\mathrm{m}}}
\frac{c^3\, r}{\left( r^{2}+c^{2}\right)
(c^{2}+(r-1)^{2})(c^{2}+(r+1)^{2})}<1\ .
\label{rmenor}
\end{equation}

Cooling occurs when the ratio is negative since then the  radiation pressure is negative. Recall that $c>0$ and then the negative-pressure condition is
simply%
\[
r<0
\]%
which implies the simple condition%
\beq
\frac{1}{\Omega _{\mathrm{m}}}\left( G\overline{x}+\Delta \right) <0\ . \label{cond}
\eeq

Let us now consider the implications of the bound \eqref{rmenor}.
A close examination of the function $f(r)$ shows that the maximum value occurs for $c=c_{\rm max}=1/\sqrt{2}$ at $r=r_{\rm max}=c$,
where
\beq
f(r_{\rm max})\bigg|_{c_{\rm max}} =\frac{G\big|\alpha _{\mathrm{max}}\big|^{2}}{4\pi \Omega _{%
\mathrm{m}}}<1\ .
\eeq
This condition therefore ensures stability for all values of
$c$. On the other hand, for a given $c$ the stability condition on the parameter $G\big|\alpha _{\mathrm{max}}\big|^{2}/ \Omega _{%
\mathrm{m}}$ may be weaker.

%The behavior of $f(r)$ depends on the parameter $c$.
%There is a critical value $c_{\rm cr}\approx 0.2135 $
%below which the function $f(r)$ has two maxima and one minima, i.e. for $c<c_{\rm cr}$. On the other hand, for $c\geq c_{\rm cr}$ the function has only one maximum.

\begin{figure}[h!]
 \centering
 \includegraphics[width=0.46\textwidth]{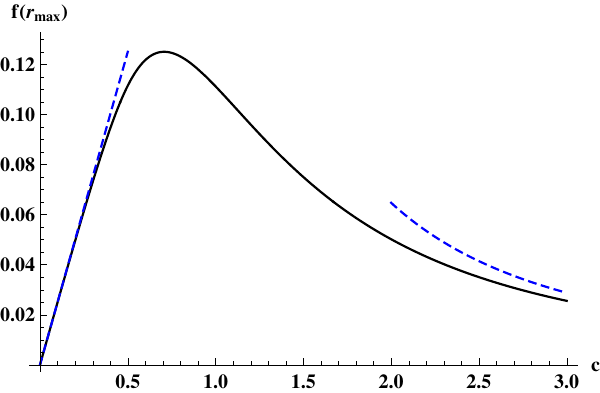}
  \caption
 {The maximum value of the ratio $f(r)\equiv
\frac{\left\langle F\dot{x}\right\rangle}{P_{fric}}$ 
as a function of $c$ (in units of $2G\big|\alpha _{\mathrm{max}}\big|^{2}/\pi \Omega _{%
\mathrm{m}}$). The dashed lines correspond to the small $c$ and large $c$ behavior, governed by the formulas
\eqref{ccsma} and \eqref{hyt}, respectively.
 }
 \label{ffrm}
 \end{figure}

\medskip

For example, for small values of $c$ the absolute maximum is located
at $r^2_0\approx 1-2c^2$. At the maximum, the bound becomes
\beq
f(r_0)\bigg|_{c\ll 1}=\frac{G\big|\alpha _{\mathrm{max}}\big|^{2}}{2\pi \Omega _{%
\mathrm{m}}}\ c < 1\ .
\label{ccsma}
\eeq
This reproduces the small $c$ behavior of fig. \ref{ffrm}.

For  $c\gg 1$, the function $f(r)$ has  the maximum located at $r_1\approx c^2/5+O(1)$, and the stability condition becomes
\beq
f(r_1)\bigg|_{c\gg 1}=\frac{25\sqrt{2}}{108}\, \frac{G\big|\alpha _{\mathrm{max}}\big|^{2}}{\pi \Omega _{%
\mathrm{m}}}\ \frac{1}{c^2} < 1\ ,
\label{hyt}
\eeq
in agreement with fig. \ref{ffrm}.
As $c\gg1$, this implies a weaker  condition on the parameters
$G\big|\alpha _{\mathrm{max}}\big|^{2}/ \Omega _{%
\mathrm{m}}$.
Indeed we have checked that the condition \eqref{ccsma} arising at small values of $c$ is
sufficient to ensure stability at all values of $c$
(at least in the small amplitude regime assumed in this subsection).

The equation \eqref{cond} obtained implies that for
blue-detuning $\Delta >0$ one could also have cooling is the mechanical
off-set $\overline{x}$ is such that 
$G\overline{x}<-\Delta $. If we discuss, as in \cite{ludwig2008optomechanical}, in terms purely of the detuning $\Delta$ with the off-set $\overline{x}=0$, then the condition is simply that red-detuning \footnote{For red-detuning $\Delta <0$ one can typically have cooling. To be more precise, the theory of resonator dynamics indicates that in this regime the energy tends to be transferred from the resonator to the cavity \cite{marquardt2006dynamical}. The back-action damping is positive, guaranteeing that the system will remain in a state where the mechanical oscillator fluctuates about a fixed point.} leads to cooling whereas blue-detuning to amplification, as usual. \\

%%%%%%%%%%%%%%%%%%%%%%%%%%%%%%

\section{Limit cycles}

The same mathematical considerations can be applied to the study of limit cycles in \cite{lorch2014laser}. An effective equation of motion, in the form of a Fokker-Planck (FP) equation was derived using a general parametrized phase-space quasiprobability distribution, which can be reduced to the P representation, Q (Husimi) representation, or the Wigner function $W$. The FP equation in \cite{lorch2014laser} is transformed into polar coordinates, and the focus is on the time evolution of the oscillator amplitude $r$ by integrating out the phase variable $\phi$ \footnote{The drift and diffusion coefficients of the Fokker-Planck equation are independent of the phase of the mechanical phase-space variable $\beta$ due to the rotating-wave approximation used in \cite{lorch2014laser}.}. This results in a one-dimensional Fokker-Planck equation for the $Q$-function of the mechanical oscillator
\beq
\label{uno}
\dot Q(r)=-\partial_r \mu(r) Q(r)+\partial_r^2 D(r) Q(r)\ ,
\eeq
where $\mu (r)$ and $D(r)$ are the  drift and diffusion coefficients. All this is explained in more detail in Appendix A, summarizing the basic formalism developed in \cite{lorch2014laser}.  It is worth mentioning that, while in the main text in \cite{lorch2014laser} the case of the $Q$-function was considered, the more general parametrized phase-space quasiprobability distribution was presented as well and the drift coefficient mantains the same form regardless of the initial quasiprobability distribution used to describe the system's density operator. However, the expression for the diffusion coefficient depends on the phase-space distribution. This can be seen in the expression given for the case when the phase-space distribution is the Wigner function \cite{lorch2014laser}. This more complex expression for the diffusion coefficient can also be fully evaluated using Bessel summation formulas, as we demonstrate explicitly below for diffusion corresponding to the Wigner function.

%However, the diffusion coefficient is specific to the phase-space distribution \cite{lorch2014laser}, but it is equally amenable to our Bessel summation formulas, as we show below for the case of the Wigner function. **** Ya he modificado esta sugerencia del refereee 2.

\subsection{Drift and diffusion coefficients of the effective equation of motion}

We start by providing an exact calculation of the infinite sums describing the drift and diffusion coefficients in the Fokker-Planck formulation of \cite{lorch2014laser}. We explain the origin, providing also context for these expressions in Appendix A (where the notation and conventions are also further discussed). They are given by \cite{lorch2014laser}
\bea
\mu(r)&=& -\gamma r+g_0E^2 \sum_n
{\rm Im}\left[\frac{J_{n-1}(\eta r)J_{n}(\eta r)}{h_{n-1}h_n^*}\right]\ ,
\label{muorig}
\\
D(r)&=&\frac12\gamma(\bar n+1)+\frac{g_0^2E^2}{2}\sum_n\left(
\frac{\kappa J_n(\eta r)^2}{|h_n|^2|\tilde h_{n-1}|^2}-{\rm Re}\left[\frac{J_{n-2}(\eta r) J_n(\eta r)}{\tilde h_{n-1}h^*_{n-2} h_n}\right]\right)\ ,
\label{ddff}
\eea
where
\beq
h_n=\kappa+i(n\omega_m-\Delta_{\rm eff})\ ,\qquad \tilde 
h_n=\kappa+i(n\omega_m-\tilde \Delta_{\rm eff})\ .
\eeq
Here $r$ represents the amplitude, $E$ is the laser drive rate, $g_0$ is the single-photon coupling strength, the dimensionless coupling $\eta=2g_0/\omega_m$ is used to characterize the optomechanical coupling \cite{lorch2014laser}, $\kappa$ is the cavity photon-number (energy) decay rate (linewidth); the field amplitude decays at $\kappa/2$ and
$\Delta=\omega_L-\omega_c$ is the detuning between the laser frequency $\omega_L$
and the bare cavity resonance $\omega_c$. The mechanical damping rate parameter is denoted by $\gamma$. Note that we denoted the mechanical frequency by $\Omega_{m}$ in the first part of the paper and we now use $\omega_{m}$.\\

The Kerr effect causes a nonlinear, intensity-dependent
shift in resonance frequency that is straightforward for the mean field
but more complex for fluctuations due to quantum effects. The difference
between these two effective detunings arises because the mean field
experiences a static shift proportional to mean photon number, while
fluctuations experience additional dynamical corrections reflecting
their nonclassical nature.\\

The term effective detuning $\Delta_{\rm eff}$ refers to the fact
that the cavity detuning is dynamically modified due to the presence of
Kerr nonlinearity and depends on the intracavity amplitude. This means
that the resonance of the cavity is shifted depending on the photon
number inside the cavity, which leads to a nonlinear self-consistency
for the detuning. When analyzing the system at the level of the mean field (the coherent amplitude $\alpha$), the Kerr nonlinearity induces a shift in the cavity resonance
  proportional to the mean photon number, so the effective detuning for
  the mean field is \cite{lorch2014laser}
  \[
\Delta _{\rm eff}=\Delta +2K\sum_{n}\left\vert \alpha _{n}\right\vert ^{2},
\]
where $K$ denotes the strength of an effective Kerr nonlinearity with $K=\frac{g_{0}^{2}}{\omega _{m}}$ and  \(|\alpha|^2\) is the mean photon
number. This accounts for the shift seen by the average field in the cavity.

For the quantum fluctuations around the mean field, the situation is more subtle.
Fluctuations ``see'' a slightly different detuning because their dynamics are affected by both the mean field shift and the changing photon number, which can be time-dependent and has its own variance. The
expression for the effective detuning for fluctuations must take into account higher-order corrections and the time dependence in the Kerr interaction. This leads to
\[
\tilde \Delta _{\rm eff}=\Delta +4K\sum_{n}\left\vert \alpha _{n}\right\vert ^{2}.
\]
The distinction between the effective detuning for the mean
field and that for the fluctuations arises from the different
ways the Kerr nonlinearity affects these two components of the optical
field. The difference can often be neglected if the fluctuations are small compared to the mean field. This is typically the case in the regime where the system is well away from bistability or instability thresholds, and photon number fluctuations are much smaller than the mean photon number (i.e., coherent state or weak
nonlinearity). These two expressions for the effective detunings are derived in \cite{lorch2014laser} with their semipolaron formalism.\\

The static approximation for the detuning is valid for a negligible Kerr parameter $K$ or in the weak driving limit. Indeed, in the $%
\left\vert \alpha (t)\right\vert ^{2}\ll 1$ limit the bare detuning is just shifted by the constant Kerr term  and there is no need to distinguish between $\Delta_{\rm eff}$ and $\widetilde{\Delta }_{\rm eff}$. The resulting Kerr shift of the detuning $\Delta _{\rm eff}=\Delta +K$ in that limit is also derived from the semipolaron transformation in \cite{lorch2014laser}, not as a phenomenological correction. It comes from how the radiation pressure interaction effectively displaces the mechanical oscillator based on the cavity photon number. This displacement, when properly accounted for, feeds back into the cavity frequency, leading to that renormalized detuning.
This reflects a self-consistent frequency shift of the cavity due to the presence of photons — the hallmark of the optical Kerr effect, here induced via radiation pressure. Previous work \cite{armour2012quantum} had noticed the shift in numerical studies and in \cite{nation2013nonclassical} it had been noted that a renormalized detuning is essential for capturing nonclassical limit-cycle behavior.\\
%appears naturally in
%the framework of \cite{lorch2014laser} (see also \cite{armour2012quantum,nation2013nonclassical}) and explains the appearance of limit cycles for specific particular values given in \cite{lorch2014laser} and also, for example, for limit cycles on the blue sideband in \cite{nation2013nonclassical}. So, already in the static approximation model, this classical Kerr shift is of importance.\\ 

A consideration of both the static and the dynamical setting is alternatively presented throughout the whole discussion in Ref. \cite{lorch2014laser}. Here, we first discuss the static case, valid for $\left\vert \alpha (t)\right\vert ^{2}\ll 1$ and then in Section B.1 below, we discuss the dynamical case by evaluating $\Delta_{\rm eff}(r)$ in full, in contrast to the sideband-resolved approximation of \cite{lorch2014laser}. Indeed, these sums have been evaluated in a close-to-resonance approximation \cite{lorch2014laser}, where one discards an infinite number of terms. Using the summation formulas of appendix B, we can now provide exact expressions for the complete resummation.
For the drift, we obtain 
\beq
\label{muyy}
\mu(r)=-\gamma r- \frac{\pi g_0E^2}{\omega_m}{\rm Re}\left[\frac{1}{2\kappa-i\omega_m}\left(\frac{J_{\nu^*}(\eta r)J_{1-\nu^*}(\eta r)}{\sin(\pi\nu^*)}+\frac{J_{1+\nu}(\eta r)J_{-\nu}(\eta r)}{\sin(\pi\nu)}\right)\right]\ ,
\eeq
with
\beq
\label{nup}
\nu\equiv \frac{\Delta_{\rm eff}}{\omega_m}+\frac{i\kappa}{\omega_m}\ .
\eeq
Details of the calculation are given in  appendix B and below we will discuss its asymptotic value for large $\eta r$.
For the diffusion coefficient, we find
\beq
\label{duyy}
D(r)=\frac12 \gamma(\bar n+1)+D_a(r)+D_b(r)\ ,
\eeq
where
\bea
&& D_a(r)=-\frac{\pi}{2\omega_m}\, g_0^2E^2{\rm Im}\left[ B\left(\frac{J_{\nu}(\eta r)J_{-\nu}(\eta r)}{\sin(\pi\nu)}
+\frac{J_{1+\tilde \nu^*}(\eta r)J_{-1-\tilde\nu^*}(\eta r)}{\sin(\pi\tilde \nu^*)}\right) \right]
\\
&& D_b(r)=\frac{\pi}{2\omega_m}\, g_0^2E^2{\rm Im}\left[ C_1\frac{J_{\nu}(\eta r)J_{2-\nu}(\eta r)}{\sin(\pi\nu)}
+C_2\frac{J_{1+\tilde \nu^*}(\eta r)J_{1-\tilde\nu^*}(\eta r)}{\sin(\pi\tilde \nu^*)}+C_3
\frac{J_{2+\nu^*}(\eta r)J_{-\nu^*}(\eta r)}{\sin(\pi\nu^*)} \right]
\eea
 $\tilde\nu = \frac{\tilde \Delta_{\rm eff}}{\omega_m}+\frac{i\kappa}{\omega_m}$ and the constants
 $B$, $C_1$, $C_2$, $C_3$ are given in appendix B.
 
\medskip

\subsubsection{Diffusion in the Wigner case}

The Fokker-Planck equation \eqref{uno} can also be written for the Wigner function, instead of for $Q$ \cite{lorch2014laser}. It has the same drift coefficient, and a diffusion given by
(see (B.20) in \cite{lorch2014laser})
\beq
\label{sumadd}
D_W=\frac14 \gamma (2\bar n+1)+\frac14  g_0^2E^2\sum_n \frac{\kappa}{|\tilde h_{n+1}|^2}\left( 
\frac{J_{n+2}J_{n+2}}{h_{n+2}h_{n+2}^*}+\frac{J_{n}J_{n}}{h_{n}h_{n}^*}-
\frac{J_{n}J_{n+2}}{h_{n}h_{n+2}^*}-
\frac{J_{n}J_{n+2}}{h_{n}^*h_{n+2}}\right).
\eeq
All the arguments of the Bessel functions are $\eta r$. This expression plays a relevant role in the dynamics of limit cycles.
The infinite sums can again be computed exactly by using the summation formulas of
Appendix B.
The final result is a long expression and we will not need it here, but
$D_W$ drastically simplifies when $\tilde \Delta_{\rm eff} =\Delta_{\rm eff}$.
This  identification is justified in the sideband-resolved regime studied in \cite{lorch2014laser}.
We obtain (see Appendix B for details)
\beq
\label{dtodo}
D_W =\frac14 \gamma (2\bar n+1)+\frac14\, g_0^2E^2\left( D_W^{(1)}+D_W^{(2)}+D_W^{(3)}+D_W^{(3)*}\right)\ ,
\eeq
where
\bea
 && D_W^{(1)}=-\frac{\pi}{\omega_m^2} 
 {\rm Im}\left[\frac{J_{-\nu}J_\nu}{\beta \, \sin(\pi\nu)}-\frac{J_{\nu+1}J_{-\nu-1}}
 {\beta^* \, \sin(\pi\nu)}\right]\ ,\qquad \beta=\omega_m -2i\kappa\ ,
\nonumber
\\ \nonumber
\\
 && D_W^{(2)}=-\frac{\pi}{\omega_m^2} 
 {\rm Im}\left[\frac{J_{-\nu}J_\nu}{\beta^* \, \sin(\pi\nu)}-\frac{J_{\nu-1}J_{-\nu+1}}
 {\beta \, \sin(\pi\nu)}\right]\ ,
\nonumber
\\\nonumber
\\
 && D_W^{(3)}= 
 \frac{\pi}{2\omega_m^2(\omega_m+2i\kappa)}
 \left[ \frac{\kappa}{\omega_m+i\kappa} 
 \left(\frac{J_{\nu+2}J_{-\nu}}{\sin(\pi\nu)}-\frac{J_{\nu^*}J_{-\nu^*+2}}{\sin(\pi\nu^*)}\right)
 -\frac{iJ_{\nu+1}J_{-\nu+1}}{\sin(\pi\nu)}+\frac{iJ_{\nu^*+1}J_{-\nu^*+1}} {\sin(\pi\nu^*)}
 \right]\ ,
 %{\beta \, \sin(\pi\nu)}\right]\ ,\qquad \beta=\omega -2i\kappa
\nonumber
\eea
where $\nu $ has been defined in \eqref{nup}.

\begin{figure}[h!]
 \centering
 \begin{tabular}{cc}
\includegraphics[width=0.4\textwidth]{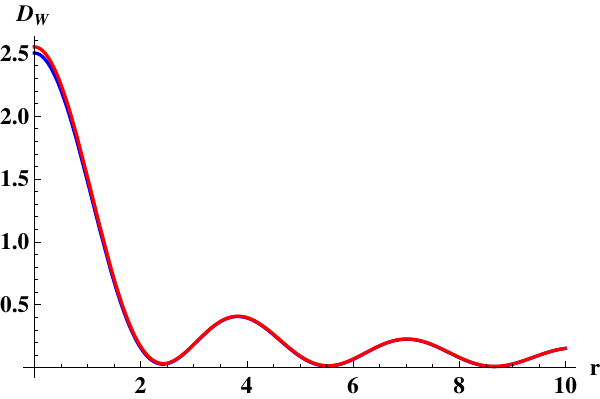}
 &
 \qquad \includegraphics[width=0.4\textwidth]{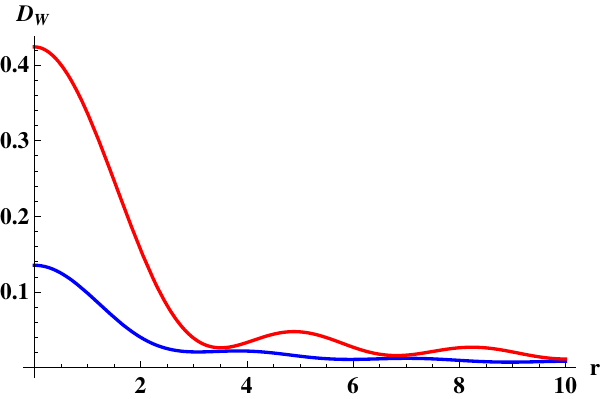}
 \\ (a)&(b)
 \end{tabular}
 \caption
 {Optical part of the Wigner diffusion from approximation \eqref{ccoo} (blue) and
 for the exact formula \eqref{dtodo} (red) for $\kappa=0.1\omega_m $. a) $\Delta_{\rm eff}=0.1\, \omega_m$.  The approximation \eqref{ccoo} reproduces  with high accuracy the exact curve. b) $\Delta_{\rm eff}=0.6\, \omega_m$. The approximation
 \eqref{ccoo} is no longer applicable as higher $n$ terms in the infinite sum \eqref{sumadd} become important.}
 \label{ddab}
 \end{figure}

\begin{figure}[h!]
 \centering
 \includegraphics[width=0.46\textwidth]{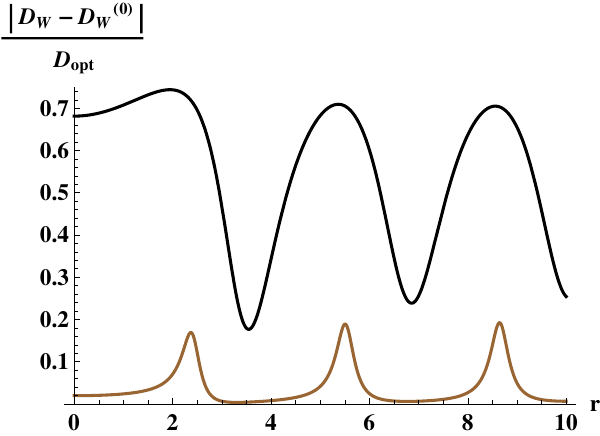}
  \caption
 {Fractional differences as a function of $r$. For  $\Delta_{\rm eff}=0.1\, \omega_m$ (brown curve) the difference reaches up to 20 per cent, whereas for  $\Delta_{\rm eff}=0.6\, \omega_m$ (black curve) the difference is over 70 per cent for some values of $r$.
 }
 \label{f2new}
 \end{figure}

These exact formulas may be compared with the approximation in Equation (48) of \cite{lorch2014laser}, valid
close to the resonance point $n=0$. This approximation requires $\Delta_{\rm eff}\ll \omega_m$ in which case the original sum \eqref{sumadd} reduces to
the $n=0$ term,
\beq
\label{ccoo}
D_W ^{(0)}=\frac14 \gamma (2\bar n+1)+\frac{\kappa g_0^2E^2}{\omega_m^4}\left(J_1^2(\eta r)
+\frac{\omega_m^2}{2(\kappa^2+\Delta_{\rm eff}^2)}\, J_0^2(\eta r)\right)\ .
\eeq
In figs. \ref{ddab}, \ref{f2new}, we compare our exact formula for $ D_W$ with $D_W^{(0)}$
for $\kappa=0.1$, $\omega_m$ (resolved sideband) and two different values of $\Delta_{\rm eff}$.
We see that the approximation \eqref{ccoo} is very inaccurate  for $\Delta_{\rm eff}=0.6 \, \omega_m$, with a difference that oscillates between 20 and 75 percent (see fig. \ref{f2new}). 
For $\Delta_{\rm eff}=0.1 \, \omega_m$, the difference can reach 20 per cent for some values of $r$.
%A closer examination shows that the approximation 
%\eqref{ccoo} is no longer applicable (with an error higher than 20 per cent) 
%for $\Delta_{\rm eff} > 0.4\, \omega_m$.
%Fractional differences

On the other hand, the exact representation \eqref{dtodo} holds for any value of the parameters, and therefore can be applied even for very high values of $\Delta_{\rm eff}$ and $\kappa$.
For large values of $r$, one can use the asymptotic formulas for the Bessel
functions and express the full result in terms of trigonometric functions.
Examples of this procedure are given in the next section.

\subsection{Asymptotic formulas, classical optomechanics cycles and the dynamical setting.}

\begin{figure}[h!]
 \centering
 \begin{tabular}{ccc}
\includegraphics[width=0.3\textwidth]{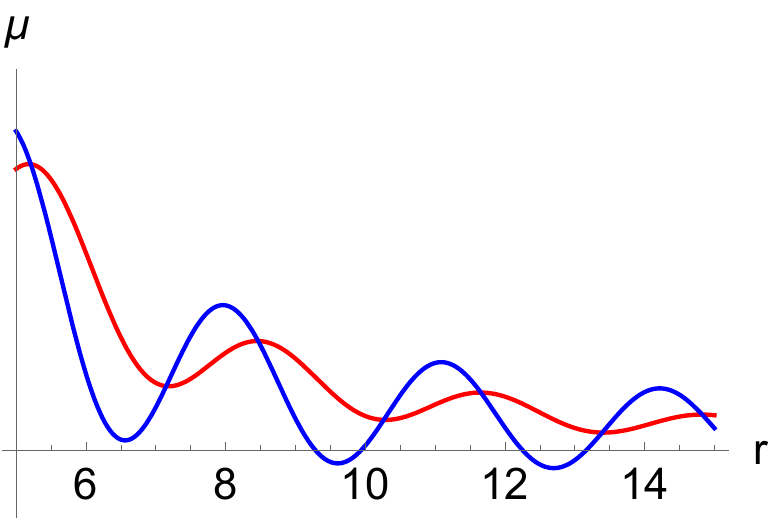}
 &
 \quad \includegraphics[width=0.3\textwidth]{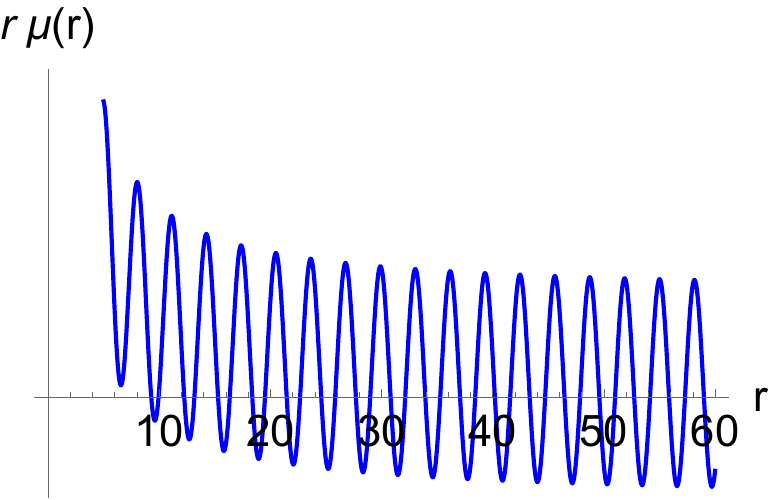}
  &
 \quad \includegraphics[width=0.3\textwidth]{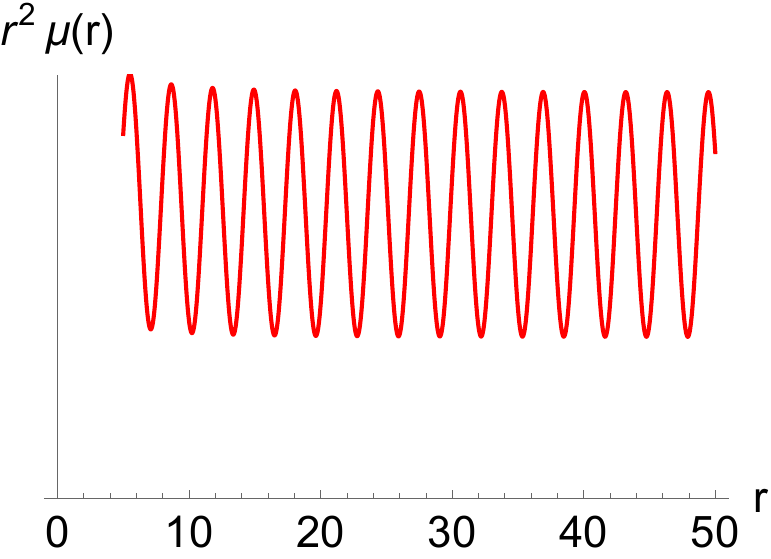}
 \\ (a)&(b)&(c)
 \end{tabular}
  \caption
 { The exact $\mu(r)$ for two different values of $\Delta$.
 Oscillations are  suppressed by an extra factor $1/r$
when $\Delta=n \omega_m$. Here $\gamma=0$, $\kappa=\omega_m=1$.
(a) 
In blue, $\Delta=0.8\omega_m$, in red $\Delta=\omega_m$. (b) Plot of $r\mu(r) $
for $\Delta=0.8 \omega_m$, exhibiting the $\mu\sim 1/r$ asymptotic behavior  when  $\Delta/\omega_m$ is not an integer. (c) Plot of $r^2\mu(r) $
 exhibiting the $\mu\sim 1/r^2$ behavior  for  $\Delta = \omega_m$.
 }
 \label{muaa}
 \end{figure}

In order to see the implications of these formulas, it is useful to look at the asymptotic behavior for large arguments of the Bessel function, in a parameter regime where the full resummation is important. That is, for frequencies away from resonance: in this regime all  secondary resonances significantly contribute to the sum. For large $x$, the Bessel functions have the familiar asymptotic expression \cite{watson1922treatise}
\beq
\label{asymJ}
J_\nu(x)\approx \frac{\sqrt{2}}{\sqrt{\pi x}}\, \cos(x-\nu\frac{\pi}{2}-\frac{\pi}{4})\ .
\eeq
Using this formula, we can find a revealing expression for the  drift \eqref{muyy} in the regime of large $\eta r$.
We obtain

\beq
\label{muasy}
\mu(r)=-\gamma  r-\frac{8 E^2 g_0 \kappa  \sin \left(\frac{\pi  \Delta }{\omega_m}\right)
   \cosh \left(\frac{\pi  \kappa }{\omega_m}\right) \cos (2 \eta 
   r)}{\eta  r \omega_m  \left(4 \kappa ^2+\omega_m ^2\right) \left(\cosh \left(\frac{2 \pi 
   \kappa }{\omega_m }\right)-\cos
   \left(\frac{2 \pi  \Delta }{\omega_m }\right)\right)}\ .
\eeq
To simplify the notation, here we write $\Delta\equiv\Delta_{\rm eff}(r)$. This formula exhibits the  periods of oscillations in $r$ and in $\Delta/\omega_m$.
A surprising effect predicted by the asymptotic formula for $\mu(r)$ is that
the second term in \eqref{muasy} identically vanishes for
\beq
\Delta =n\omega_m, \qquad n\in \mathbb{Z}\ .
\eeq
%More precisely, the leading term in the asymptotic expansion at large $r$ vanishes.
Thus, for these special values of $\Delta$, $\mu(r) =-\gamma r+O(1/r)$.
Oscillations are suppressed by a factor  $1/r$
originating from subleading terms in the asymptotic expansion of the Bessel functions. This effect is shown in fig. \ref{muaa} and is a result of the full resummation. It is not seen if one
truncates the sum to few values of $n$, as in a close-to-resonance approximation. 
%It would be interesting to see this effect in experiments.

An important challenge is to observe this effect experimentally. This requires designing the cavity such that the detuning, $\Delta=\omega_L-\omega_c$,
 equals an integer multiple of the mechanical frequency, 
$\omega_m$. In this case, measuring the drift 
$\mu(r)$ as a function of amplitude should reveal the behavior shown in Fig. \ref{muaa}c: the product 
$r^2\mu $  oscillates, with its maxima approaching a constant value for large amplitudes
$r$. This behavior is clearly distinguishable from the case 
$\Delta\neq n \omega_m$, where the maxima of 
$r^2\mu $ increase linearly with the amplitude.

\medskip

\subsubsection{Classical optomechanical limit cycles and a comparison with close to resonance results}
The theory for classical optomechanical limit cycles, as presented in \cite{lorch2014laser}, is described by just the drift component \eqref{muyy} of the Fokker-Planck equation. In the absence of diffusion and with a constant effective detuning $\Delta_{\rm eff}$, the evolution of the oscillator amplitude is deterministic, and the oscillator amplitude satisfies the following equation \cite{lorch2014laser} 
\beq
\dot r=\mu(r) =-r\ \gamma_{\rm eff}(r)\ ,\qquad \gamma_{\rm eff}(r)=\gamma+\gamma_{\rm opt}(r)\ .
\eeq
The combined intrinsic and optically induced damping of the oscillator is denoted by $\gamma_{\rm eff}(r)$. The amplitudes $r_0$ for limit cycles are determined
by the conditions $\gamma_{\rm eff}(r_0)=0$
and $\gamma_{\rm eff}'(r_0)>0$. Taking into account the complete sum defining $\mu(r)$
derived above in \eqref{muyy}, we get an explicit formula for $\gamma_{\rm opt}(r)$
which therefore allows us to  accurately determine the limit cycles.

This can be compared with the approximation used in 
(43) of \cite{lorch2014laser} for the oscillator close to resonance,
\beq
\gamma_{\rm opt}^{(0)}(r)=-\frac{g_0^2E^2}{\omega_m^2}\ \frac{2\kappa \Delta_{\rm eff}}{\Delta_{\rm eff}^2+\kappa^2}\, \frac{1}{r}\, J_0(\eta r)J_1(\eta r)\ .
\label{aproL}
\eeq

\begin{figure}[h!]
 \centering
 \begin{tabular}{cc}
\includegraphics[width=0.4\textwidth]{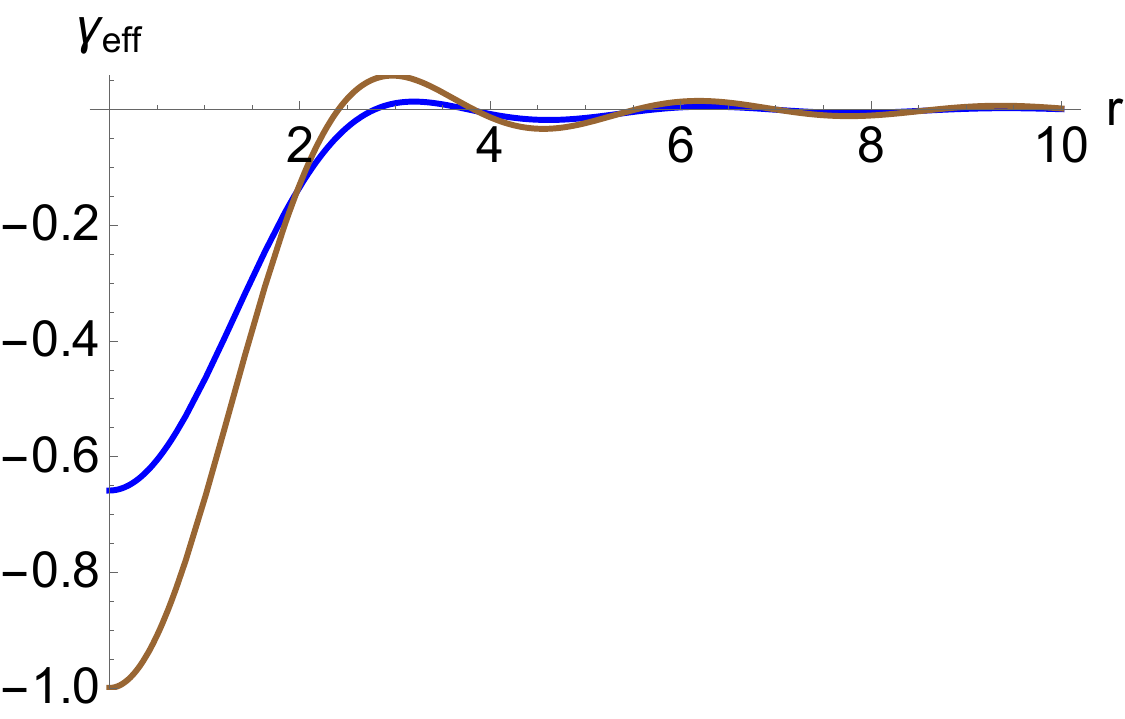}
 &
 \qquad \includegraphics[width=0.4\textwidth]{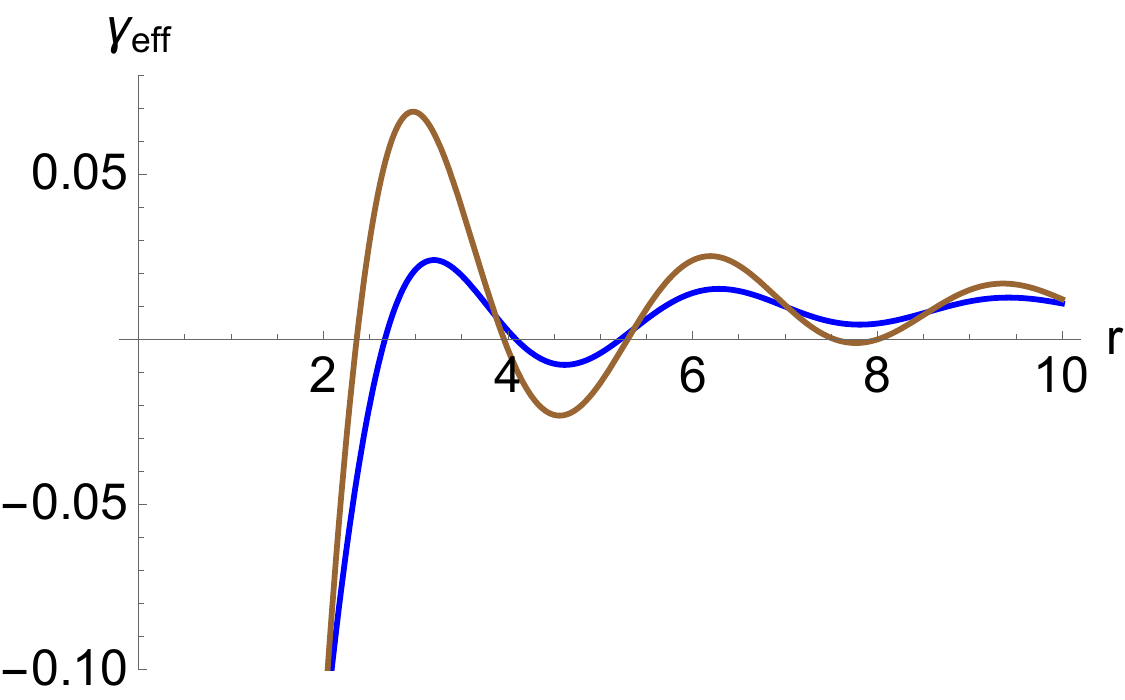}
 \\ (a)&(b)
 \end{tabular}
 \caption
 { Effective damping $\gamma_{\rm eff}$ as a function of the amplitude
 in units of $\gamma_0=g_0E^2/\omega_m^2$. Limit cycles corresponds to the zeros. 
 (a) $\gamma=0$. The approximation \eqref{aproL} (brown) compared with the $\gamma_{\rm eff}$ obtained from the complete sum (blue). We see that the location of the zeros is significantly corrected when all the terms of the sum are included.
(b) $\gamma=0.01 \gamma_0$. In this case the approximated $\gamma$ (in brown) exhibits   three limit cycles occurring at $r\approx 2.4$, $r\approx 5.3$ and $r\approx 8.1$, but the last limit cycle disappears in  
the more accurate description using \eqref{muyy} (in blue).
}
 \label{figab}
 \end{figure}

\noindent In contrast, $\gamma_{\rm opt}(r)$ with $\mu $ given by \eqref{muyy}  keeps the contributions of all values of $n$, not only $n=0$ and $n=1$.
The resulting plot has a similar shape. However,
for some values of the parameters there are crucial differences,
since some limit cycles that  appear in the approximation
\eqref{aproL} turn out to be  an artifact of the approximation. Indeed, they disappear as solutions in the  exact formula. This is shown in fig. \ref{figab}, done for
$\Delta_{\rm eff}=\kappa$, $\kappa =0.6\, \omega_m$.
(cf. with fig. 1a) of \cite{lorch2014laser}).
Although for these values of parameters there are significant modifications in the limit cycles, in a sideband-resolved
regime with $\Delta_{\rm eff},\ \kappa\ll \omega_m$ the differences obviously become smaller. Nonetheless, this example shows that the complete sum could be useful
in a more accurate description of the limit cycles in a  much wider range of parameters.

\medskip

\subsubsection{Dynamical setting}

We can similarly make use of the summation formula to compute a formula for $\Delta_{\rm eff}(r)$ itself, in the dynamical setting.
This is determined by a nonlinear equation \cite{lorch2014laser}
\beq
\Delta_{\rm eff}=\Delta+2KE^2\sum_n \frac{J_n(r\eta) J_n(r\eta )}{|\kappa+i(n\omega_m-\Delta_{\rm eff})|^2}\ .
\eeq
The infinite sum can again be computed using the summation formulas of Appendix B by noting that this can be re-writen as
\beq
\Delta_{\rm eff}=\Delta+\frac{i KE^2}{\kappa\omega_m} \sum_n \left(\frac{J_n(r\eta) J_n(r\eta )}{n-\nu}-\frac{J_n(r\eta) J_n(r\eta )}{n-\nu^*}\right)\ .
\eeq
with $\nu =\frac{\Delta_{\rm eff}}{\omega_m}+i\frac{\kappa}{\omega_m} $, as before \eqref{nup}.
Now  the summation formula gives
\beq
\Delta_{\rm eff}=\Delta+\frac{2\pi KE^2}{\kappa\omega_m} {\rm Im}\left[ \frac{J_\nu(r\eta) J_{-\nu}(r\eta )}{\sin(\pi\nu)}\right]\ .
\eeq
One can similarly study the asymptotic behavior at large $r$. In this regime $\Delta_{\rm eff}$ becomes
\beq
\Delta_{\rm eff}\approx \Delta-\frac{4 KE^2}{\kappa\omega_m\eta r} \, 
\frac{\sinh(\pi \frac{\kappa}{\omega_m})}{\cosh(2\pi \frac{\kappa}{\omega_m})-\cos(2\pi \frac{\Delta_{\rm eff}}{\omega_m})}
\left(\cosh(\pi \frac{\kappa}{\omega_m})+\cos(\pi \frac{\Delta_{\rm eff}}{\omega_m})\sin(2r\eta) \right)\ .
\label{unre}
\eeq
This can be solved numerically, or analytically when $\Delta_{\rm eff}\ll \omega_m$. In this case
\beq
\Delta_{\rm eff}\approx \Delta-\frac{2 KE^2}{\kappa\omega_m\eta r} \, 
\frac{1}{\sinh(\pi \frac{\kappa}{\omega_m})}
\left(\cosh(\pi \frac{\kappa}{\omega_m})+\sin(2r\eta) \right)\ .
\label{unresolved}
\eeq
This formula encodes the information of all secondary resonances.
Notice that  \eqref{unresolved} is also valid in the opposite regime of the unresolved sideband (bad cavity limit) and therefore it provides a good analytical evaluation for $\Delta_{\rm eff}$ for $c=\kappa /2\omega _{m}>1$ too,  a regime not covered in \cite{lorch2014laser}. This regime is of great physical interest because it is easily achieved as the mechanical oscillator size increases, since larger oscillators are characterised by lower frequencies.

%It evaluates $\Delta_{\rm eff}$ not only for $\left\vert \alpha (t)\right\vert ^{2}\gg 1$, when the static approximation is invalid, but  also in the unresolved sideband limit

%\begin{figure}[h!]
% \centering
% \includegraphics[width=0.46\textwidth]{figextraMaxDif.pdf}
%  \caption
% {The fractional difference between the exact 
% $\Delta_{\rm eff}$ obtained by the summation formula and  $\Delta_{\rm eff}^{(0)}$ obtained by truncating the sum including the main resonance and the first secondary resonance (here $\nu=0.5+i 0.3$),
% }
% \label{fexnew}
% \end{figure}

 \smallskip
 It is instructive to compare the 
$\Delta_{\rm eff}$ 
 obtained from the exact summation formula with that derived from an approximation based on retaining the main resonance. While the difference depends on the specific parameter values, it is possible to anticipate the conditions under which this discrepancy becomes significant.
If $\kappa\ll \omega_m$ and $\Delta/\omega_m$ is very close to a resonance, i.e. very close to an integer $\Delta \approx n_0\omega_m$, then the term $n=n_0$
dominates over all other terms. Any other $n$-term in the sum is suppressed by a factor $\frac{\kappa^2}{\omega_m^2}\frac{1}{(n-n_0)^2}\ll 1$. In this case, retaining the main resonance provides a good approximation.\\
%In concrete experimental setups, e.g. as in \cite{lorch2014laser} (taking the values of parameters of fig. 2 in this paper),
%such difference between can reach up to 9 per cent e.g. for $\eta r=2.4$, but is generally around 2-5 per cent.
However, the most important deviations between the exact and the approximate formula arise when $\Delta_{\rm eff}$ is not close to a resonance. Let us consider an example.
If $KE^2$ is small, one has $\Delta_{\rm eff}\sim \Delta$ in a leading-order approximation. 
If $\Delta/\omega_m$ is a half integer, then oscillations in the amplitude $r$ are suppressed in the exact formula, as can be seen from \eqref{unresolved} by virtue of the fact that $\cos(\pi\frac{\Delta_{\rm eff}}{\omega_m})=0$. However, oscillations are not suppressed in a truncation that keeps only one or two terms of the sum. This generally leads to a significative difference in the numerical results. The difference
can also be especially relevant, in particular,  when it implies a change of sign in the total $\Delta_{\rm eff}$ with regards to the sign of $\Delta$ \cite{nation2013nonclassical,lorch2014laser} (recall the discussion of the static approximation in Sec III.A).
%This is an interesting physical open problem that stems from these formulas.\\

In addition to these numerical differences, it is important to note that an approximation based on a truncated sum fails to capture certain properties that emerge only upon full resummation over $n$ -- specifically, those arising at special values of $r$ and $\Delta_{\rm eff}$, for which the trigonometric functions in \eqref{unre} may vanish.

%Recall the previous discussion in Section III.A regarding the static approximation of detuning. Even in the static case, where the detuning is shifted by $K$, this modification can be crucial if the detuning is negative without correction but becomes positive with $K$. This implies that whereas negative detuning involves cooling, it may allow amplification and the existence of a limit cycle \cite{nation2013nonclassical, lorch2014laser}. Therefore, a significant discrepancy between the static approximation and the sideband approximation can be especially relevant for the dynamics if the resolved approximation correction does not change the sign of a red detuning, whereas the full correction we evaluated does change it to a positive effective detuning.\\

%%%%%%%%%%%%%%%%%%%%%%%%%%%%%%%
\section{Outlook}

We have seen that the radiation pressure of a monochromatic light source can be fully analysed. This extends the analytical understanding of such a central physical quantity \cite{aspelmeyer2014cavity,marquardt2006dynamical,marquardt2007quantum} and allows a number of consequent considerations, both at the exact level and for asymptotic analysis for regions of parameter space that would otherwise be inaccessible. We have seen how the physics can then be analysed for the whole parameter space without the usual limitations which are due to the lack of evaluation of all the sideband contributions.\\

Indeed, the type of expressions we have evaluated, which in addition to the radiation pressure includes also the drift and diffusion coefficients, including Wigner diffusion, in the Fokker-Planck formulation of the effective equations of motion \cite{lorch2014laser}, are known to be quite ubiquitous in cavity optomechanics and in other problems of the Floquet type \cite{russo2024landau,tierz2025analytical}. Therefore, many other uses of the evaluations here can be devised, both for work along the lines we have considered here, such as \cite{amitai2017synchronization,armour2012quantum,rodrigues2010amplitude}, and for other settings and/or quantities not explored here but manifestly amenable to a similar or even analogous treatment. The set of examples includes the consideration of periodic modulation of optical resonators \cite{minkov2017exact,crespo2023optimized} and the study of scattering of N-photons in a periodically modulated cavity \cite{trivedi2020analytic}, optomechanical systems with the inclusion of thermal effects \cite{hu2021generation}, the generation of frequency combs both at the optic and at the magnonic level \cite{hu2021generation,crespo2023optimized,xu2023magnonic,gu2024optical,wang2024optomechanical}, the study of Floquet-Weyl semimetals \cite{wang2022nonlinear}, the use of analogies between the coupled sideband dynamics of optomechanical oscillations and a Bloch-band structure \cite{ding2025bloch} or the study of photon-photon correlation function \cite{qian2012quantum}. Recently, there has been considerable interest and progress in extending quantum amplification features by considering a Floquet setting \cite{jiang2022floquet,jiang2021floquet,su2022review}. The analytical evaluations presented here (and in related studies \cite{russo2024landau,tierz2025analytical}) can also be applied to further study spin Floquet amplification.\\

%It is manifest from a cursory examination of these references that 
% each of these research lines would benefit directly from the application of the summation formula and its subsequent 
%asymptotics. 
It is evident that the implementation of the summation formula, along with its exact properties and asymptotics, would prove to be of significant benefit to each of these research lines.
%A selection of these applications will be reported elsewhere. 
On the other hand, the feasibility of applying these methods to other problems, involving synchronization or to the multiple drive setting, such as those outlined in \cite{holmes2012synchronization,wood2021josephson} for example, appears to be a distinct possibility in principle but it is important to note that the analytical study for example of equilibrium conditions in \cite{holmes2012synchronization} or the generalized Bessel setting of \cite{wood2021josephson}, is more arduous.\\

%The technical advance is this paper is not only the existence of a compact simple formula describing all terms in the summation, but the resulting applicability of asymptotic expressions gives expressions in terms of uniquely elementary functions. 
The technical advance of this paper lies in the formulation of a compact, simple formula that comprehensively encompasses all terms within the summation. This provides an analytical understanding of the parameter space and offers insight into the underlying physics of the various processes.
Additionally, it facilitates the application of asymptotic expressions, resulting in expressions that are expressed in terms of uniquely elementary functions.
Since these are simple trigonometric functions, further elucidation, for example of the zeros \cite{tierz2025analytical}, becomes a particularly straightforward task, helping to identify special points in regions of the parameter space in limits where a resolved sideband approximation, which retains only a few resonant terms in the summation, does not reach. There are also open problems already in the context of the present discussion, for example, whether the more complete analysis of Wigner diffusion presented here also allows a more detailed analysis of energy fluctuations by evaluating the Fano factor, as in \cite{lorch2014laser}, but with full summation. All these considerations require no further mathematical developments than those presented here or in the mathematically related works \cite{russo2024landau,tierz2025analytical}.\\

A novel extension, such as the consideration of multichromatic instead of monochromatic light, would imply the substitution of Bessel functions by generalised Bessel functions \cite{korsch2006two,lang2023discrete}. Progress along these lines is possible and will be reported elsewhere, even though the fundamental summation formula does not extend naively to this setting. The interest in this extended setting goes beyond optomechanics, magnomechanics or optical resonators applications as it is also a fundamental extension in a number of quantum Floquet problems with periodic modulations \cite{geng2022self,minkov2017exact,wood2021josephson}, including the consideration of a number of quantum technologies platforms where having multiple drives is essential for the control of the quantum system.
%, for example. 

\vspace{12pt} 
\begin{acknowledgments}
We thank Professors David Pérez-García, Oriol Romero-Isart, Patrick Maurer, Wenjie Ma and Jiteng Sheng for discussions and/or correspondence at different stages of this work. The work of MT was started at Universidad Complutense de Madrid (UCM) and continued and finished after moving to the Shanghai Institute for Mathematics and Interdisciplinary Sciences (SIMIS). This work was
supported by the Shanghai Institute for Mathematics and
Interdisciplinary Sciences (SIMIS) under the Applied and
Interdisciplinary Collaboration Project. M.T. gratefully acknowledges the resources and facilities provided by SIMIS, which were essential for the completion of this research. JGR acknowledges financial support from grant 2021-SGR-249 (Generalitat de Catalunya) and  by the Spanish  MCIN/AEI/10.13039/501100011033 grant PID2022-126224NB-C21. MT acknowledges previous financial support from FEI-EU-22-06, funded by Universidad Complutense de Madrid, and grant PID2020-113523GB-I00, funded by the Spanish Ministry of Science and Innovation.

\end{acknowledgments}

\newpage
\appendix

\section{Fokker-Planck description of limit cycles}

The paper \cite{lorch2014laser} develops an analytical Fokker-Planck formalism for describing
mechanical limit cycles in the quantum regime of optomechanical systems. After summarizing the conventions and notation used through the main text, we present a summary of this formalism.

\subsection*{Conventions and notation}

Drive and energy. We reserve $E$ \emph{only} for the \emph{laser drive rate} (units of s$^{-1}$), sometimes also called drive amplitude, related to the input field by $E=\sqrt{\kappa_{\rm ex}}\,\alpha_{\rm in}$ and $P_{\rm in}=\hbar\omega_L|\alpha_{\rm in}|^2$. This is not to be confused with the \emph{mechanical energy}, denoted $\mathcal{E}=\hbar\omega_m r^2$. We use $K=g_0^2/\omega_m$ and $\eta=2g_0/\omega_m$.\\

Detuning and effective detunings. $\Delta=\omega_L-\omega_c$. We use $\Delta_{\rm eff}=\Delta+2K\sum_n|\alpha_n|^2$ for the optical mean field and $\tilde\Delta_{\rm eff}=\Delta+4K\sum_n|\alpha_n|^2$ for fluctuations (twice the shift). In the sideband-resolved regime studied in \cite{lorch2014laser}, these coincide to leading order.\\

Damping-rate conventions (factor-of-two). There are two conventions as explained in \cite{lorch2014laser}
(i) The one used in \cite{lorch2014laser} and hence in our second part: $\kappa$ denotes the \emph{field-amplitude} decay rate in the Heisenberg--Langevin equation, $\dot a=(i\Delta-\kappa)a+\sqrt{2\kappa}\,a_{\rm in}$; the photon-number (energy) decay rate is $2\kappa$. (ii) The one in the earlier works \cite{marquardt2006dynamical,ludwig2008optomechanical} (and hence used in our Part I): $\kappa_{\rm(energy)}$ denotes the photon-number decay rate and the field amplitude decays at $\kappa_{\rm(energy)}/2$. Therefore a factor of 2 between the two. We adhere to these respective conventions within each part.\\

\textbf{Starting Point: Optomechanical Master Equation}\\

The dynamics of the coupled cavity-mechanical system is governed by the
standard quantum master equation \cite{aspelmeyer2014cavity}:

\[
\frac{d}{dt} \rho = (\mathcal{L}_m + \mathcal{L}_c + \mathcal{L}_\text{int})\rho,
\]

where:

\begin{itemize}
%\tightlist
\item
  \(\mathcal{L}_m\): Liouvillian for the mechanical oscillator,
\item
  \(\mathcal{L}_c\): Liouvillian for the cavity,
\item
  \(\mathcal{L}_\text{int}\): Cavity-mechanics interaction term.
\end{itemize}

Explicitly, these are:

%\[
\begin{align*}
\mathcal{L}_m\rho &= -i[\omega_m b^\dagger b, \rho] + \gamma(\bar{n} + 1)\mathcal{D}[b]\rho + \gamma\bar{n}\mathcal{D}[b^\dagger]\rho, \\
\mathcal{L}_c\rho &= -i[-\Delta a^\dagger a - iE(a-a^\dagger), \rho] + \kappa\mathcal{D}[a]\rho, \\
\mathcal{L}_\text{int}\rho &= -i[-g_0 a^\dagger a (b + b^\dagger), \rho],
\end{align*}
%\]
where \(a, b\) are photon and phonon (mechanical) annihilation
operators, respectively, and \(g_0\) is the single-photon optomechanical
coupling. The frequency of the mechanical oscillator is $\omega _{m}$, its amplitude damping rate is $\gamma =\frac{\omega _{m}}{Q_{m}}$, and the mean phonon number in
thermal equilibrium is $\overline{n}$. The Lindblad operators are denoted by $\mathcal{D}[a]$ \cite{lorch2014laser} and $\kappa$ is the cavity photon-number (energy) decay rate (linewidth); the field amplitude decays at $\kappa/2$ and $\Delta $  is the detuning from cavity resonance of the driving field $E$.\\

\textbf{2. Semipolaron Transformation and Phase-Space Representation}\\

The formalism uses a \textbf{phase-space representation} (such as the Husimi Q-function) for the mechanical oscillator, introducing a joint phase-space operator \(\sigma(\beta,\beta^*)\). Through a
``semipolaron'' transformation the optical Kerr
nonlinearity is separated and the nonlinearity of the optomechanical
interaction can be considered in an analytically tractable way. This transformation is designed to facilitate the analysis of nonlinear quantum dynamics in cavity optomechanical systems, especially
when studying limit cycles beyond simple linearized theories.\\

This results in a hybrid description where the mechanical
  oscillator is described in phase space while the optical cavity
  remains in Hilbert space.
%\item
  The main feature of the semipolaron transformation is to partially ``dress'' the cavity field, introducing a phase
  shift that depends on the mechanical oscillator's position in phase space. The transformation is formally defined as:
%\end{itemize}
\[
\widetilde{\sigma}(\beta,\beta^*, t) = e^{\eta(\beta-\beta^*)a^\dagger a/2} \sigma(\beta,\beta^*, t) e^{-\eta(\beta-\beta^*)a^\dagger a/2}\ ,
\]
or equivalently as a unitary transformation:
\[
\widetilde{\sigma}(\beta, \beta^*, t) = e^{i\theta(\beta,\beta^*)a^\dagger a} \sigma(\beta,\beta^*, t) e^{-i\theta(\beta,\beta^*)a^\dagger a}\ ,
\]
with \(\theta(\beta, \beta^*) = \eta \operatorname{Im}(\beta)\) and
\(\eta = 2g_0/\omega_m\). This approach explicitly separates the cavity's Kerr nonlinearity (arising
from the radiation pressure coupling) from the optomechanical
interaction, while still retaining a residual interaction term.
Importantly, it preserves a clear interpretation of the mechanical state
by allowing for a well-defined reduced density operator for the
mechanics.\\

This transformation is named ``semipolaron'' because it bears similarity to the full polaron transformation (which entirely diagonalizes the interaction at
the cost of entangling cavity and mechanics), but only ``partially'' transforms: it keeps the two subsystems largely separate, allowing calculations in which second-order perturbation theory can be applied to the remaining interaction term.\\

In contrast, the standard polaron transformation would remove the
optomechanical interaction completely but prevent straightforward
access to the reduced state for the mechanical oscillator, which is
essential when studying quantum limit cycles. The Liouvillian for the cavity acquires an effective Kerr nonlinearity, and the drive becomes phase-modulated in a way that depends on the mechanical oscillator.\\

\textbf{3. Adiabatic Elimination and Effective Fokker-Planck Equation}\\

  Now the jump operator for cavity dissipation remains unchanged---unlike in
  a full polaron picture---while the crucial relation between
  \(\widetilde{\sigma}(\beta, \beta^*, t)\) and the mechanical
  Q-function is preserved. This separation enabled the authors in \cite{lorch2014laser} to perform an adiabatic elimination of the cavity. By adiabatically eliminating the cavity and focusing on the mechanical
degree of freedom, the effective Fokker-Planck equation (FPE) for the
reduced Q-function (or equivalently, the Wigner function, with suitable
corrections) of the mechanical oscillator is obtained. Schematically,
\[
\frac{\partial}{\partial t} Q(\beta, \beta^*) = \text{Drift} + \text{Diffusion} + \text{Coupling terms}.
\]
After phase averaging and in polar coordinates
(\(\beta = r e^{i\varphi}\)), the FPE reduces to a 1-dimensional form
for the amplitude \(r\):
\[
\frac{\partial}{\partial t} Q(r) = -\partial_r [\mu(r) Q(r)] + \partial_r^2 [D(r) Q(r)],
\]
where the key quantities (corresponding to \eqref{muorig} and \eqref{ddff} in the main text) are:\\
\newline
\textbf{Drift coefficient:}

\[
\mu(r) = -\gamma r + \sum_n g_0 E^2 \, \mathrm{Im}\left[ \frac{J_{n-1}(\eta r) J_n(\eta r)}{h_{n-1} h^*_n} \right]
\]
\\
\textbf{Diffusion coefficient:}

\[
D(r) = \frac{\gamma(\bar{n}+1)}{2}
+ \sum_n \frac{g_0^2 E^2}{2} \left[
\frac{\kappa J_n^2(\eta r)}{|h_n|^2 |\tilde{h}_{n-1}|^2}
- \mathrm{Re}\left( \frac{J_{n-2}(\eta r) J_n(\eta r)}{\tilde{h}_{n-1} h^*_{n-2} h_n} \right)
\right]
\]

with

\begin{itemize}
%\tightlist
\item
  \(\kappa\): cavity photon-number (energy) decay rate (linewidth); the field amplitude decays at $\kappa/2$,
\item
  \(J_n\): Bessel functions (from sidebands generated by mechanical
  motion),
\item
  \(h_n = \kappa + i (n\omega_m - \Delta_\text{eff})\),
\item
  \(E\): laser drive rate,
\item
  \(\Delta_\text{eff}\): effective detuning including Kerr effect.
\end{itemize}

This formalism captures nonlinear limit-cycle dynamics in the quantum regime. This Fokker-Planck approach bridges optomechanical laser theory with quantum statistical mechanics, enabling tractable yet powerful predictions for quantum effects in experimentally relevant parameter regimes. In this work we have evaluated analytically both the drift and diffusion coefficient and the ensuing asymptotic behavior in different regimes and its consequences, thereby extending the analytical results of the Fokker-Planck formalism of \cite{lorch2014laser}.

\section{Bessel summations and diffusion coefficients}

In order to compute the infinite sums 
in the drift and diffusion coefficient expressions in \eqref{ddff}, we will need the following summation formulas \cite{russo2024landau,newberger1982new}:
\bea
\sum_n \frac{J_{n-1}(x)J_n(x)}{n+\mu} &=& -\frac{\pi}{\sin(\pi\mu)} \, J_{-\mu}(x)J_{\mu+1}(x)
\nonumber\\
\sum_n \frac{J_{n}(x)J_n(x)}{n+\mu} &=& \frac{\pi}{\sin(\pi\mu)} \, J_{-\mu}(x)J_{\mu}(x)
\nonumber\\
\sum_n \frac{J_{n-2}(x)J_n(x)}{n+\mu} &=& \frac{\pi}{\sin(\pi\mu)} \, J_{-\mu}(x)J_{\mu+2}(x)
\label{sumaf}
\eea
We first write the denominator in \eqref{muorig} as follows
\beq
\frac{1}{h_{n-1}h^*_n}=\frac{1}{2\kappa-i\omega_m}\left( \frac{1}{h_{n-1}}+\frac {1}{h^*_{n}}\right)\ .
\eeq
Substituting this expression into \eqref{muorig},
a direct application of the summation formulas \eqref{sumaf} yields the compact analytic expression \eqref{muyy}.

\subsection{Diffusion coefficient}

Now, in addition to the formulas above, it is convenient to write the expression
as sum of terms of with simpler denominators linear
in $n$.
We have
\bea
&& \frac{1}{|h_n|^2|\tilde {h}_{n-1}|^2}=\frac{B}{2\kappa }\left(\frac{1}{h_{n}}+\frac {1}{\tilde {h}^*_{n-1}}\right)+\frac{B^*}{2\kappa }
\left(\frac{1}{h_{n}^*}+\frac {1}{\tilde {h}_{n-1}}\right)
\\
&& \frac{1}{h_n{\tilde {h}}_{n-1}h^*_{n-2}}=\frac{C_1}{h_{n}}-\frac {C_2}{\tilde {h}_{n-1}}-\frac{C_3}{h_{n-2}^*}
\eea
with
\beq
B\equiv \frac{1}{(\omega_m+\tilde \Delta_{\rm eff}- \Delta_{\rm eff})(\omega_m+\tilde \Delta_{\rm eff}- \Delta_{\rm eff}-2i\kappa)}
\eeq
\beq
C_1\equiv \frac{1}{2(\omega_m+\tilde \Delta_{\rm eff}- \Delta_{\rm eff})(\omega_m-i\kappa)}\ ,\quad
C_2\equiv  \frac{1}{(\omega_m+\tilde \Delta_{\rm eff}- \Delta_{\rm eff})(\omega_m-\tilde \Delta_{\rm eff}+ \Delta_{\rm eff}-2i\kappa)}\ ,
\eeq
\beq
C_3\equiv  \frac{1}{2(\omega_m-i\kappa)(\omega_m-\tilde \Delta_{\rm eff}+ \Delta_{\rm eff}-2i\kappa)}
\eeq
Now the formulas \eqref{sumaf} can be directly applied.
In the particular case $\tilde \Delta_{\rm eff}=\Delta $, this gives the closed expression \eqref{dtodo}.

\subsection{Wigner diffusion}

Once again the sums can be computed exactly using the  summation formulas \eqref{sumaf}.
As in the previous cases, the strategy is to write each fraction as sum of individual terms containing
$1/h_n$ or $1/\tilde h_n$. For example, for the first term in the diffusion expression, for the case of Wigner diffusion, given by \eqref{sumadd}, we have
\beq
\frac{\kappa}{|\tilde h_{n+1}|^2| h_{n+2}|^2}=\frac{1}{2(\omega-\Delta+\tilde \Delta)}\left(\frac{A}{h_{n+2}}+\frac{A^*}{h_{n+2}^*}+
\frac{A^*}{\tilde h_{n+1}}+\frac{A}{\tilde h_{n+1}^*}\right)\ ,
\eeq
with
$$
A\equiv \frac{1}{\omega -\Delta+\tilde\Delta-2i\kappa}\ ,
$$
and similarly for the rest of the terms.

\bibliography{cavityBIB2}

 \end{document}